\documentclass[prb,aps,twocolumn,showpacs]{revtex4}
\usepackage{epsfig}
\begin{document}
\title{Structural transitions in the 309-atom magic number Lennard-Jones cluster}
\author{Eva G. Noya}
\author{Jonathan P. K. Doye} 
\affiliation{University Chemical Laboratory, Lensfield Road, Cambridge CB2 1EW,
	United Kingdom}
\date{\today}

\begin{abstract}

	The thermal behaviour of the 309-atom Lennard-Jones cluster, whose
structure is a complete Mackay icosahedron, has been
studied by parallel tempering Monte Carlo simulations. Surprisingly for
a magic number cluster, the heat 
capacity shows a very pronounced peak before melting, which is attributed to 
several coincident structural transformation processes. The main transformation is 
somewhat akin to surface roughening, and involves a cooperative condensation of vacancies 
and adatoms that leads to the formation of pits and islands one or two layers 
thick on the Mackay icosahedron. The second transition in order of importance
involves a whole scale transformation of the cluster structure, and
leads to a diverse set of twinned structures that
are assemblies of face-centred-cubic tetrahedra with 6 atoms along their edges,
i.e., one atom more than the edges of 
the 20 tetrahedra that make up the 309-atom Mackay icosahedron. A surface 
reconstruction of the icosahedron from a Mackay to an anti-Mackay overlayer is also observed, 
but with a lower probability.

\end{abstract}

\pacs{61.46.Bc,36.40.Ei,36.40.Mr}

\maketitle
\vspace{0.5cm}
\section{Introduction}

	Lennard-Jones (LJ) clusters have been widely studied over
the last few decades and have become a model system for understanding
some of the structural, thermodynamic and dynamic properties that
are particular to clusters. This choice is mainly due to the simplicity of the
model, but it also reflects the possibility of making 
comparisons with experiments on rare gas clusters. 
For example, the
ground-state structure of all LJ clusters with up to 1600 atoms has
now been 
investigated.\cite{northby,pillardy,jon95,jonchem95,deaven,wales,romero,%
hartke,leary,xiang,xiang-2,xiang-3,note1} 
It is well-established that, for $N < 1000$, 
LJ clusters follow an icosahedral
pattern growth, showing magic numbers corresponding to
complete Mackay icosahedra at $N=$ 13, 55, 147, 309, 561, and 923.
In between these magic numbers, the vast majority of the global
minima correspond to a Mackay icosahedron core covered by an
incomplete outer layer. The exceptions occur at or near to shell 
closings of alternative structural forms. Mostly, these
correspond to Marks decahedra, but there is also one instance
of a face-centred-cubic (fcc) truncated octahedron ($N=38$)\cite{pillardy,jon95} and of a Leary 
tetrahedron ($N=98$).\cite{leary} 

\begin{figure}
\begin{center}
\includegraphics[angle=270,width=85mm]{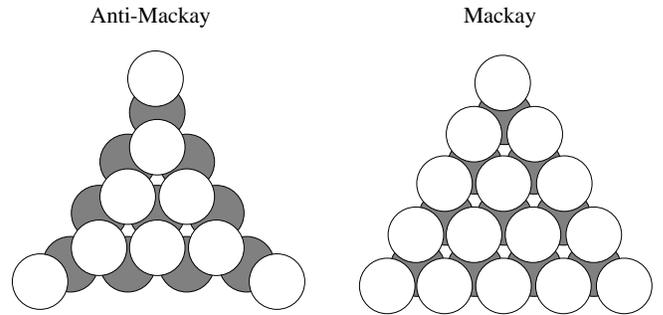}
\caption{\label{fig1} Atomic positions for the two possible
overlayers on a 147-atom Mackay icosahedron.}
\end{center}
\end{figure}

	There are two different ways in which a layer can grow on the surface
of a Mackay icosahedron, which are termed Mackay
and anti-Mackay growth modes.\cite{northby} A Mackay icosahedron
can be considered to be made up of twenty slightly distorted 
fcc tetrahedra sharing a common vertex. 
The Mackay growth mode continues the
fcc packing of the underlying tetrahedra,
leading to the next Mackay layer, whilst the anti-Mackay growth
mode involves sites that are in hexagonal close packing 
positions with respect to the underlying tetrahedra, thus 
introducing a twin plane. The
two overlayers are illustrated in Fig.\,\ref{fig1}.
The growth of a new shell in LJ clusters usually starts at 
the anti-Mackay sites, as this overlayer does not involve 
low-coordinate edge atoms.
However, as the  Mackay overlayer is associated with a higher
surface density, it has a lower strain energy and 
becomes energetically favourable over the anti-Mackay above a given
size.

	The thermal behaviour of LJ clusters has also been 
much studied.\cite{berry,labastie,cheng,frantz-0,calvo,jon,neirotti,%
florent,flo01,frantz,jon2,jon4,sabo,predescu,liu,mandelshtam,florent2}
It is known that many clusters do not
retain the ground-state structure up to melting, but, instead,
they undergo structural changes as the temperature is raised.
For LJ clusters, 
solid-solid transitions prior to melting can be expected
for those sizes with non icosahedral 
geometries.\cite{jon,neirotti,florent,jon4,sabo,predescu,liu,florent2}
Premelting effects associated with structural changes at the
surface that lead to enhanced surface diffusion are also
common.\cite{cheng,calvo} Of particular
interest to us here are the surface reconstructions which
can occur for clusters with an incomplete Mackay overlayer and
in which the outer layer changes
from Mackay to anti-Mackay.\cite{jon2,mandelshtam} This surface reconstruction
is mainly driven by the greater vibrational entropy associated 
with anti-Mackay structures. For a particular shell,
the transition is first seen at the crossover size at which 
a Mackay structure becomes lowest in energy (e.g.
at $N=31$ for the second shell, and at $N=82$ for the
third shell), and also occurs at  
the subsequent sizes but at a temperature that increases with the
cluster's size. Therefore, there comes a size for which the melting temperature
becomes lower than the temperature necessary for the surface reconstruction
to occur and, so, the surface reconstruction only takes place in a 
limited size range, namely
$N=31-39$\cite{frantz,jon2,mandelshtam} for the
second Mackay layer and $N=82-140$\cite{mandelshtam} 
for the third  Mackay layer. 

	As the growth of the fourth shell of the Mackay icosahedron 
follows the same pattern
as the previous ones, it is very likely
that the surface reconstruction again occurs.
The results for the second and third layer seem to indicate that 
there might be a tendency for the size range over which 
the surface reconstruction takes place to increase with the number
of shells. In particular, if the upper size for which this transition
occurs continues to increase with the shell number, it might be 
that even the complete
309-atom Mackay icosahedron could undergo such a surface 
reconstruction as well. This would be an unexpected result, since 
usually magic numbers clusters will show a single `first-order-like' 
melting peak in the heat capacity,
and `premelting' and other transitions are more generally
associated with incomplete shells. Indeed, Labastie and Whetten's classical
paper on the size evolution of cluster melting
towards the bulk first-order behaviour, was based on the Mackay 
icosahedral LJ clusters with 13, 55 and 147 atoms.\cite{labastie}
However, there is some precedent for such a transition in 
a complete Mackay icosahedron. For
Morse clusters, it has already been shown that such a 
surface reconstruction occurs prior to melting
for $N=$ 561 and 923,\cite{jon3} although the parameter in the Morse
potential that determines the range of the potential
had a value for which anti-Mackay structures are
stabilized relative to the LJ potential.\cite{jon5}
Therefore, a study of the melting behaviour of the 309-atom
LJ cluster would serve both to test our hypothesis concerning
a possible surface reconstruction and to better understand
the thermodynamics of surface transitions in clusters.

\section{Methods}

	The potential energy of a cluster interacting with the
LJ potential is given by
\begin{equation}
E = 4 \epsilon \sum_{i<j} \left[ \left(\frac{\sigma }{r_{ij}}\right) ^{12} 
- \left(\frac{\sigma }{r_{ij}}\right) ^{6}\right],
\end{equation}
	where $\epsilon $ is the pair well depth and $2^{1/6}\sigma $
is the equilibrium pair separation.
	
	The thermal behaviour of the LJ 309-atom complete Mackay
icosahedron was studied using canonical Monte Carlo (MC) simulations
with the parallel tempering exchange scheme (PT).\cite{pt1} The
PT MC simulations consisted of 32 trajectories at temperatures 
ranging from 0.360\,$\epsilon /k_B$ to 0.453\,$\epsilon /k_B$. 
A preliminary study over a broader range of temperatures, from 
0.010\,$\epsilon /k_B$ to 0.600\,$\epsilon /k_B$, showed that the
region 0.360\,$\epsilon /k_B$ - 0.453\,$\epsilon /k_B$ was wide 
enough to capture all the interesting features of the heat capacity 
curve. Due to the slow convergence, the system was equilibrated by 
performing 7.2 $\times 10^{7}$ MC cycles for each one of the 
trajectories, and the data to compute the heat capacity were 
collected over the following 3.06$\times 10^{8}$ MC cycles. Similar 
convergence problems have already been encountered in previous PT MC
studies of the thermal behaviour of smaller LJ 
clusters.\cite{neirotti,florent,sabo,predescu,liu,mandelshtam}
Exchanges between two trajectories were attempted with a probability 
of 10\%. All the temperatures were initialized with the Mackay icosahedron
ground-state structure. Evaporation and fragmentation of the cluster was
avoided by adding a hard wall of radius $r+\sigma $, where $r$ is the 
radius of the ground-state structure. The heat capacity and caloric curves
were obtained from the simulation data 
using the multihistogram technique.\cite{labastie}

\section{Results}
	
\begin{figure}
\begin{center}
\includegraphics[width=85mm]{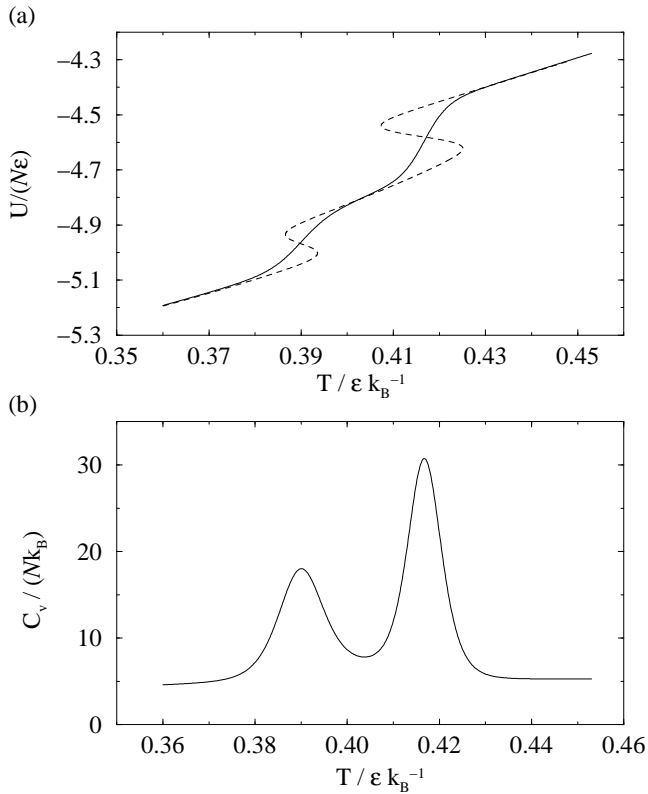}
\caption{\label{fig2} (a) Canonical (solid line) and microcanonical
	(dashed line) caloric curves and (b) canonical heat capacity 
	for the 309-atom LJ cluster.}
\end{center}
\end{figure}

	In Fig.\ \ref{fig2} we have plotted the canonical and
microcanonical caloric curves and the canonical heat capacity, as 
obtained from our simulations using the multihistogram technique.
There are two fairly sharp changes in the slope of the canonical 
caloric curve, which, in the microcanonical ensemble, lead to 
two van der Waals loops, i.e. regions in which the caloric curve
exhibits a negative slope and the heat capacity is negative. 
Coincident with these features in the caloric curves, at 
the temperatures 0.390\,$\epsilon /k_B$ and 0.417\,$\epsilon /k_B$, the 
canonical heat capacity curve shows two clearly distinct peaks. The height of 
the first peak is roughly half the height of the second one, but the former 
is slightly broader. The latent heat of these transitions was computed as 
the area under each peak in the heat capacity curve. It was found that the 
latent heat associated with the first peak is approximately $0.67\,l_{m}$, 
where $l_{m}=0.29/N\epsilon $ is the latent heat of the second peak, 
which, as we will show below, signals the melting of the cluster. These 
results indicate that the 309-atom LJ cluster undergoes a major structural 
transition before melting. The unusual presence of two van der Waals loops in 
the microcanocial caloric curve, as well as the two strong peaks in 
the canonical heat capacity, suggest that this transition might be, as with
the cluster melting transition, the finite-size analogue of a first-order 
transition.\cite{labastie} In the regions where the microcanonical heat
capacity is negative, the canonical probability distribution
for the energy is also bimodal.\cite{wales94}

	The structural transformations associated with
the two peaks in the heat capacity curve were analyzed by performing 
systematic quenches starting from configurations generated in each of 
the 32 trajectories employed in our calculations.\cite{wales90} Two useful properties 
to characterize the minima are their potential energy and geometric
mean vibrational frequency ${\bar \nu }$. The main contributions to ${\bar \nu }$
come from atoms which are nearest neighbours. For a pair potential,
such as LJ, the value of ${\bar \nu }$ reflects both the number of 
nearest neighbours, and the average nearest-neighbour distance, the
latter because the second derivative of the LJ potential decreases
with distance. Strained structures, which inevitably have a variety
of nearest-neighbour distances, have a larger average nearest-neighbour
distance, because of the anharmonic nature of the LJ well, and
hence have smaller values of ${\bar \nu }$. 

\begin{figure}
\begin{center}
\includegraphics[width=85mm]{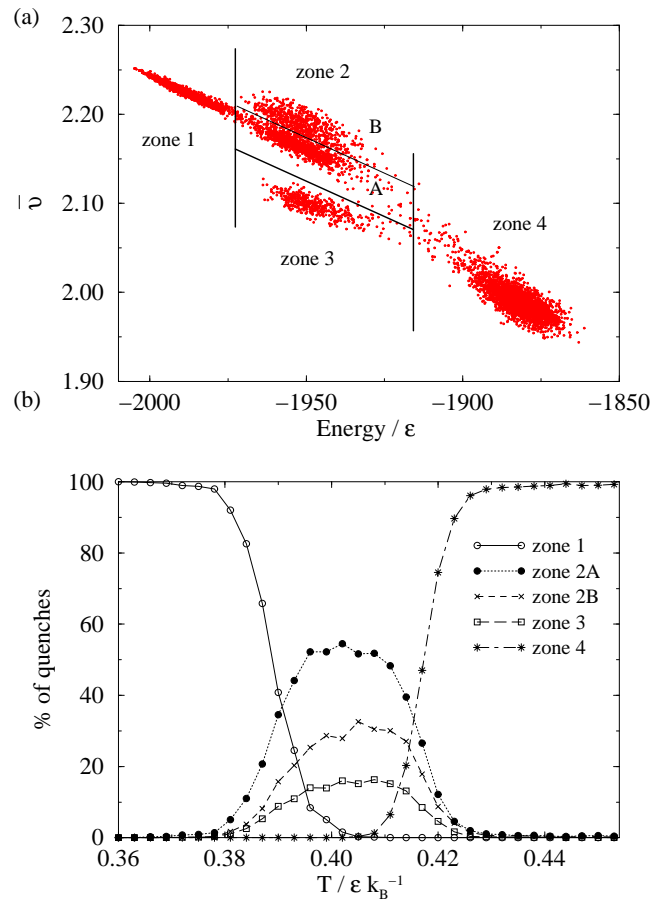}
\caption{\label{fig3} (a) (Colour online) A scatter plot of the
	geometric mean vibrational frequency against the potential
	energy for the minima collected from the quenches at all
	temperatures. Five distinct zones can be distinguished
	in this diagram, and the boundaries that we have chosen to
	delineate these zones are
	explicitly shown in the plot.
	The unit of frequency is $(\epsilon /m \sigma^2)^{(1/2)}$.
	(b) The percentage of quenches that lead to minima 
	in each of the five zones as a function of temperature.}
\end{center}
\end{figure}

	Plotting ${\bar \nu }$
against the potential energy for the minima we obtained shows that
they can be naturally divided into a number of types, which, as we 
will see, reflect their different structural features (see Fig.\,\ref{fig3}(a)). 
Specifically, the diagram 
can be divided into four well-defined zones, with only a relatively
small number of minima near to the boundaries. 
A closer inspection of zone 2 shows that this zone can be further divided
into two subzones (2A and 2B) on the basis of their frequencies, but the
division in this case is less clear cut. 

\begin{figure*}
\begin{center}
\includegraphics[width=180mm]{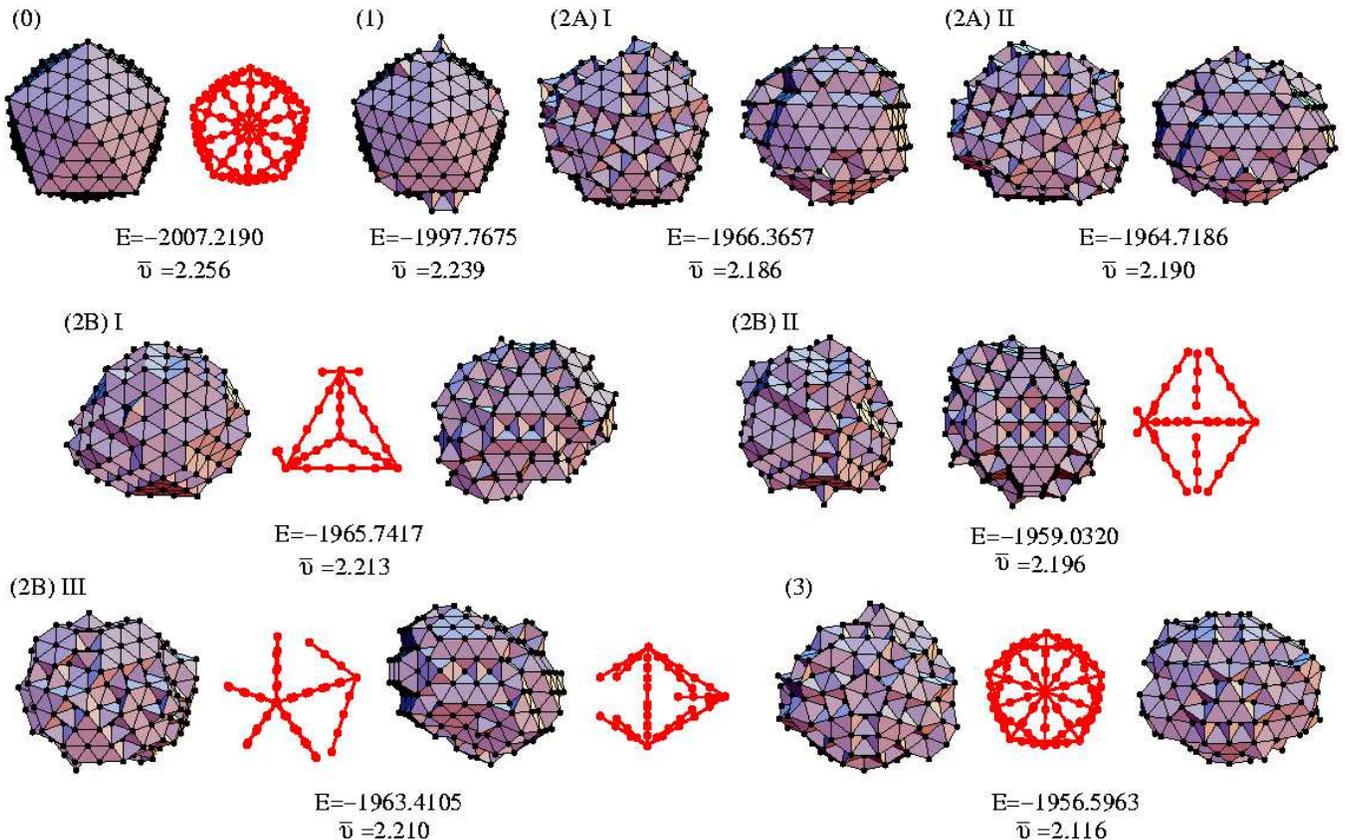}
\caption{\label{fig4} (Colour online) Structure of the global minimum and 
        of some of the isomers
	representative of the zones shown in Fig. 3(a). The energy and
	frequency are given in reduced units. To illustrate how the 
	structures can be viewed as assemblies of fcc tetrahedra, to the
	right of some of the complete structures, we have also 
	depicted the network formed by those atoms that lie along
	the edges of these tetrahedra.}
\end{center}
\end{figure*}

	To determine the structural 
types associated with the different zones in Fig.\,\ref{fig3}(a) we 
visually inspected a large number of minima. To facilitate this task,
we also performed common neighbour analyses of the 
structures,\cite{Jonsson,Clarke,Hendy01}
and so we were able to identify atoms that had a particular local
coordination environment. It was particularly noteworthy that all
the minima in zones 1, 2 and 3 are ordered in the sense that all
the atoms in the interior of these clusters have twelve nearest neighbours,
whose local coordination shell can be characterized as icosahedral,
decahedral, fcc or hcp.\cite{Hendy01} Typical isomers representative of each zone 
are depicted in Fig.\,\ref{fig4}.
	
	The lowest energy region (Energy $<  -1972\,\epsilon $), 
designated as zone 1, corresponds to the global minimum, the complete
Mackay icosahedron, and structures based upon it, in which there are some
vacancies at the low-coordinate vertex and edge sites, and the resulting adatoms 
are dispersed across the surface of the cluster, perhaps forming adatom clusters
of a few atoms (see structure (1) in Fig. \ref{fig4}). 
As the number of vacancies increases,
the potential energy of the minima increases, and their mean vibrational
frequency decreases due to the lowered number of nearest neighbours.

	At energies in the interval $-1972\,\epsilon $ to $-1915\,\epsilon$,
there are two clearly separate regions differentiated by their vibrational 
frequency. As mentioned above, the higher frequency region, 
zone 2, is further divided into two subzones designated as 2A and 
2B. The isomers found in zone 2A are Mackay 
icosahedra in which a pit is formed on one side of the cluster, whilst
an adatom island is formed on the opposite side (see structures (2A)\,I and II
in Fig.\,\ref{fig4}).
The islands prefer to occupy anti-Mackay sites, and both the 
pits and islands can be one or two layers thick. These structures are
therefore similar to those of zone 1, but have a significantly larger density
of adatoms and vacancies. Consistent with this, the zone 2A minima
form a downwardly sloping band in Fig.\,\ref{fig3}(a), that can be
seen to be an extension of the band due to the zone 1 minima. However,
there is a clear gap between these two bands at the boundaries between
zones 1 and 2A, because minima with an intermediate number of 
surface defects are less likely to be sampled.

	It used to be common to refer to a Mackay icosahedron as a 
multiply-twinned particle,\cite{marks1} 
because they are made of twenty somewhat strained
fcc tetrahedra sharing a common vertex. The twin planes correspond to
the internal faces of these tetrahedra and which are common to two
tetrahedra. For the 309-atom Mackay icosahedron there are five atoms
along the edges of the fcc tetrahedra. Other multiply twinned structures
that are also made up of aggregates of fcc tetrahedra are possible, such
as the Marks decahedra\cite{marks2} and the Leary tetrahedron.\cite{leary}

	After an extensive investigation of the structures of the minima
in zone 2B, in which the common neighbour analysis was particularly
revealing, we found that the common feature of this diverse set of 
structures was that they are aggregates of fcc tetrahedra with 6
atoms per edge. As the fcc tetrahedra are larger than for structures 
based on the 309-atom Mackay icosahedron, the structures are less
strained, and so have larger vibrational frequencies.
Three different examples of the possible arrangements of 
the fcc tetrahedra are shown in Fig.\,\ref{fig4}. 
For structure (2B)\,I there is a single complete fcc tetrahedron at the
centre of the cluster and it is related to a Leary tetrahedron.\cite{leary} Structure
(2B)\,II has two twinned fcc tetrahedra at its centre. 
Finally, structure (2B)\,III has five fcc 
tetrahedra (some incomplete) sharing a common edge, and so it has a
decahedral structure at its core. In all these structures, these
central tetrahedra are covered by overlayers that begin the
growth of new tetrahedra that are twinned with respect to the core.

	The lowest vibrational frequency region at intermediate
energies, zone 3, is occupied by isomers with a 
147-atom Mackay icosahedral core, but where the outer layer has 
undergone a surface reconstruction from a Mackay to an 
anti-Mackay overlayer. Because of the greater strain associated
with these structures, they have a lower mean vibrational
frequency. As mentioned in the introduction,
a similar surface reconstruction has been observed for some smaller 
LJ clusters with incomplete Mackay layers\cite{jon2,mandelshtam} 
and for larger complete Mackay icosahedra interacting with a 
long-ranged Morse potential.\cite{jon3} As seen in Fig.\,\ref{fig3}\,(a), 
this surface reconstruction is partly driven by the higher
vibrational entropy of the anti-Mackay overlayer. In addition,
covering the 147-atom Mackay icosahedron with a complete
anti-Mackay overlayer gives rise to a 279-atom structure,
and so there is also a configurational entropy gain associated 
with the possible arrangements of the 30 extra adatoms. 

	Note that the edges of the underlying fcc tetrahedra around which the
above anti-Mackay structures are formed have only four atoms along
their edges. Therefore, we
can conclude that the three regions at intermediate energies 
correspond to structures built around assemblies of fcc tetrahedra, but, 
depending on the length of the edges of these tetrahedra, and hence
the associated strain, the 
isomers fall in different regions of frequency. 

	Finally, the high energy region, zone 4, is associated 
with completely melted structures, in which both the core and surface 
atoms are disordered. There is a variety of coordination numbers
associated with the interior atoms in these structures. Furthermore,
the common neighbour analysis shows that the 
vast majority of the atoms do not have an ordered local coordination
shell. Therefore, zone 4 is assigned to the liquid state.

	An analysis of the quench results is presented in 
Fig.\,\ref{fig3}\,(b) showing the percentage of quenches that lead to
minima in the different zones. At temperatures below that for 
the first peak almost all isomers found correspond to Mackay 
icosahedra in which a number of vacancies and adatoms are formed 
on the cluster's surface. The vacancies mainly appear at the 
low-coordinate vertex and edge sites, whilst the adatoms diffuse 
around the surface, as seen previously for smaller Mackay icosahedra.\cite{cheng,kunz}
At temperatures coincident with the first peak 
there is a fairly sharp transition from the Mackay icosahedron with 
a small number of defects to structures which have undergone some kind 
of structural transition. Surprisingly, the Mackay to anti-Mackay 
reconstruction represents only a minor contribution to the heat 
capacity, with only up to around a 15-20\% probability of finding the cluster
with an anti-Mackay reconstructed surface. Instead, two other transitions
dominate in this range of temperatures. 

	In the main transition, which represents 
a 50-55\% of the total transition, adjacent atoms from the outer shells of the Mackay
icosahedron vacate their sites, leaving behind pits one or two 
layers thick, and the resulting adatoms form islands above the 
still intact parts of the fourth shell. As mentioned above, below 
the transition, the vacancies and adatoms associated with the 
Mackay icosahedron are dispersed across the surface and their 
number gradually increases with temperature. However, as the surface 
density of these defects increases there comes a point where there 
is a jump in the number of vacancies and adatoms, and they cluster 
together to form islands and pits. This process is driven by the
(effective) attraction between the adatoms and vacancies. Once 
a number of vacancies are formed at vertex and edge sites, the 
neighbouring atoms also lower their coordination number and, therefore, it 
becomes more favourable to remove more vertex and face atoms in 
positions adjacent to an already existing vacancy or pit. 
This transition is somewhat akin to the 
pre-roughening and roughening transtions, \cite{weeks} both of which
are known to occur for the bulk LJ \{111\} surfaces.\cite{celestini,tosatti}

	The second transition competing with the Mackay to anti-Mackay 
surface reconstruction, is not a surface transition, but involves
a transformation of the whole structure of the cluster. The resulting 
structures are multiply-twinned particles that are assemblies of fcc
tetrahedra with six atoms along each edge, 
one atom more than the tetrahedra that make up the 
309-atom Mackay icosahedron. This transition must be driven by an increase 
in the configurational entropy, as these structures have lower
vibrational entropies. Indeed, this conformational entropy is 
reflected in the diversity of structures that we observed (Fig.\,\ref{fig4}).
The probability that these structures are adopted reaches a maximum of about 30\%
midway between the two heat capacity peaks.

	Finally, as the second heat capacity peak is approached and
the cluster begins to melt, this
array of different structures becomes less probable in favour of 
disordered structures typical of the liquid state.

\section{Conclusions}

	The existence of a well-defined structural transition in the 309-atom 
LJ cluster prior to melting suggests that, as we anticipated in the
introduction of this paper, there is a tendency for the size 
range for which LJ clusters with Mackay overlayers undergo structural
transformations prior to melting to increase as the number of shells in the 
cluster increases. For the fourth icosahedral shell, we expect these
transitions to begin at around $N=169$, the size at which a Mackay overlayer 
becomes energetically more favourable than the anti-Mackay, and to
continue up to, and probably beyond, the next complete 
Mackay icosahedron at $N=309$. This compares to $N=31-39$ and
$N=82-140$ for the second and third icosahedral shells.
However, the unexpected aspect of our results was the nature of 
the transition that occurred and the size of the heat capacity peak 
that resulted. Contrary to our predictions, the Mackay to anti-Mackay 
surface reconstruction was only a minor contribution to the heat 
capacity peak. Instead, two other transitions had a more significant 
effect. In particular, the dominant contribution is from a surface 
transition involving the condensation of vacancies and adatoms that 
leads to the formation of pits and islands one or two 
layers thick on the Mackay icosahedron. Also fairly important was
another transition that leads to a complete loss of the Mackay 
icosahedral structure, and the adoption of multiply-twinned
structures based upon packings of fcc tetrahedra with six atoms 
along each edge. By contrast, the twenty strained fcc tetrahedra
that make up the 309-atom Mackay icosahedron have five atoms along 
each edge.

	The substantial latent heat, the van der Waals loop in the 
microcanonical caloric curve, and the bimodal distribution for the
energy in the canonical ensemble, all suggest that the
transition at T=0.390\,$\epsilon /k_B$ is the finite size analogue
of a first order phase transition. But does it have any analogues
in bulk systems? The two more minor transitions that contribute
to the peak clearly have no bulk analogue, as they are specific
to Mackay icosahedra and multiply-twinned particles, both of which
will not be energetically competitive in the bulk limit.
However, as we have already mentioned, the cooperative condensation
of the vacancies and adatoms to form pits and islands, respectively,
has some similarities to the bulk pre-roughening and
roughening transitions\cite{weeks} of the LJ fcc \{111\} 
surfaces.\cite{celestini,tosatti} All involve the 
loss of the uniform, flat character of the perfect \{111\} surfaces
through the formation of islands and pits, whilst retaining the
basic lattice structure. However, the pre-roughening transition
leads to a disordered flat surface, where the height fluctuations
are restricted to be of one layer thickness, whereas, for the
clusters, islands and pits are seen that can be two layers high
or deep. The surface roughening transition is defined
as the temperature at which a height-height correlation 
function diverges with distance. However, such a definition
cannot be straightforwardly applied to our cluster, because
the \{111\} facets are so small. Furthermore, the cooperative
nature of the transition is also related to the small size of 
the facets, since it allows the removal of whole faces, nucleated
by the preferential generation of vacancies at the vertices and
edges. Therefore, it would premature to say that the transition in
our clusters {\it is} the finite size analogue of either
the pre-roughening or roughening transitions, and it would be
interesting to see how the transition evolves with cluster size.

	Our results have also shown that even for
a very strong magic number, such as the 309-atom LJ cluster, the heat 
capacity curve can exhibit strong features prior to melting.
This is an important finding, because usually it is expected 
that magic numbers will show only one well-defined melting peak in the heat
capacity curve, whilst premelting effects and other structural transitions
are more generally associated with clusters with incomplete shells. 
For example, following the scaling arguments of Labastie and 
Whetten, which were based on the first three LJ Mackay 
icosahedra,\cite{labastie} it would have been expected that the 
309-atom cluster would exhibit a single heat capacity peak that
was narrower and higher than for the 147-atom cluster, hence smoothly 
converging towards the bulk limit. Our results have shown that
this an oversimplified view of the problem and that many 
surprising features can appear along the path towards the bulk limit. 
Indeed, one should expect structural transitions prior to melting
to be the norm for larger LJ clusters up to very large sizes. 
Firstly, larger Mackay icosahedra would be expected to show
similar transitions to those found here. Secondly, even at sizes 
when decahedral and then fcc structures become lowest in energy,
there are likely to be solid-solid transitions from decahedral 
to icosahedral and at larger sizes from fcc to decahedral,
driven by differences in vibrational entropy.\cite{jon4}

\acknowledgements
	The authors are grateful to the Ram\'on Areces Foundation (E.G.N.) 
and the Royal Society (J.P.K.D.) for financial support, and to Florent
Calvo for helpful discussions.

\end{document}